\documentclass[lettersize,journal]{IEEEtran}

\usepackage{amsmath,amsfonts}
\usepackage{algorithmic}
\usepackage{algorithm}
\usepackage{array}
\usepackage[caption=false,font=normalsize,labelfont=sf,textfont=sf]{subfig}
\usepackage{textcomp}
\usepackage{stfloats}
\usepackage{url}

\usepackage{verbatim}
\usepackage{graphicx}
\usepackage{cite}
\usepackage{booktabs, multirow}
\usepackage{amssymb}
\begin{document}

\title{Joint Language Modeling for In-the-Wild Speech-Scene Generation with
Distinct Token Representations}

\author{Semin Kim,~\IEEEmembership{Student Member,~IEEE,} Ju Yeon Kang,~\IEEEmembership{Student Member,~IEEE,}  Sunghwan Ahn,~\IEEEmembership{Student Member,~IEEE,} Seonuk Kim, and Nam Soo Kim,~\IEEEmembership{Senior Member,~IEEE}
\thanks{Semin Kim, Ju Yeon Kang, Sunghwan Ahn, Seonuk Kim, and
Nam Soo Kim are with the Institute of New Media and Communications,
Department of Electrical and Computer Engineering, Seoul National University, Seoul 08826, South Korea (e-mail: smkim21@hi.snu.ac.kr;
jykang@hi.snu.ac.kr; shahn@hi.snu.ac.kr; snkim@hi.snu.ac.kr;
nkim@snu.ac.kr).}}

\markboth{Journal of \LaTeX\ Class Files,~Vol.~14, No.~8, August~2021}%
{Kim \MakeLowercase{\textit{et al.}}: Joint Language Modeling for In-the-Wild Speech-Scene Generation with Distinct Token Representations}


\maketitle

\begin{abstract}
Speech-scene generation aims to synthesize intelligible target speech together with diverse environmental sounds as a coherent acoustic scene. The task is considered challenging in two respects. Speech and environmental audio differ substantially in their semantic and acoustic characteristics, yet existing approaches typically model the complete scene using a single mixed acoustic representation. At the same time, the two sounds are closely related in real-world speech scenes, whereas conventional approaches usually rely on simulated scenes mixed from independently recorded sources. We propose a large language model (LLM)-based framework for joint modeling of distinct speech and environment token streams. A source-aware neural audio codec extracts temporally aligned speech and environment tokens from mixed audio, and a pretrained LLM generates both token streams in an interleaved fashion conditioned on the natural-language scene description and speech content. This formulation preserves source-specific representations while modeling the token streams within a shared generative context. Since both streams are extracted directly from mixed recordings, the model can also be adapted to in-the-wild speech scenes without isolated sources. Our experimental results demonstrate that the proposed framework improves speech intelligibility, scene relevance, and audio quality over existing speech-scene generation techniques, with consistent gains from real-world adaptation. We further investigate the effects of distinct token representations and interleaved generation, and analyze whether the environmental sound remains stable while speech is active and how the two streams interact.
\end{abstract}

\section{Introduction}

\IEEEPARstart{R}{ecent} advances in text-to-speech (TTS) have enabled the generation of highly natural speech~\cite{tacotron2,vits,difftts,styletts2,valle,f5tts}, with fine-grained control over attributes such as speaking rate, speaker identity,  pitch, and style~\cite{fastspeech2,yourtts,naturalspeech3,makesinger}.
However, speech in real-world recordings rarely occurs in isolation. 
It is typically embedded in an acoustic scene, surrounded by environmental sounds such as wind, animals, vehicles, and crowds.
Generating speech within such a scene is a natural next step beyond conventional TTS.
We refer to this task as \textit{speech-scene generation}: given the text to be spoken (e.g., \textit{“Could I get a latte, please?”}) and a natural-language scene description (e.g., \textit{“A woman orders coffee in a café as rain falls”}), the goal is to generate a single recording in which the speech and the described environmental audio occur together as one coherent scene.

A fundamental challenge in this task arises from the different characteristics of speech and environmental audio. Speech is highly structured and constrained by linguistic content, whereas environmental audio is more diverse and often weakly structured. 
Most of the existing systems~\cite{voiceldm,voicedit,immersivetts,controlaudio} are built on text-to-audio diffusion models and generate the scene as a single acoustic representation of the mixture, in which the identities of speech and environmental audio are no longer explicitly specified. In these techniques, the generation model represents the two heterogeneous signals in a common target space, which hinders the learning of both sound components. 


However, modeling the two sounds independently is also undesirable. In a real-world speech scene, speech and the environmental audio are closely related. They are captured in the same space and share its reverberation and recording conditions. Such dependencies are largely absent in simulated scenes, where the two sounds are recorded independently and simply added together. 
Yet existing methods make only limited use of in-the-wild recordings, either fine-tuning the pretrained model while keeping the speech module frozen~\cite{voicedit} or relying on clean-source supervision that is available only in simulated scenes~\cite{immersivetts}. As a result, the dependencies of real-world speech scenes are not fully exploited.
To this end, we formulate speech-scene generation as joint language modeling over distinct speech and environment token streams. First, to keep the two sound components distinct, we develop a source-aware neural audio codec that converts a speech scene into two temporally aligned discrete token streams. Source-specific semantic distillation encourages each stream to capture the semantics of its own source, so the two streams serve as explicit generation targets instead of a single mixed representation.
Second, to model the relationship between the two sound components, a pretrained large language model~(LLM) interprets the scene description and generates the two streams in an interleaved autoregressive order. 
Each generated token is fed back into the context, so each stream is conditioned on the other throughout generation.
Third, the codec extracts both streams directly from the mixture, so the entire model can be pretrained on simulated scenes and then adapted to in-the-wild recordings without isolated sources. To reduce the gap between the two stages, we build the simulated scenes from speech segmented from in-the-wild recordings rather than clean read speech, and compose their descriptions to resemble captions of real-world recordings.

The main contributions of our work are summarized as follows:
\begin{itemize}
    \item We formulate speech-scene generation as joint language modeling over distinct speech and environment token streams. The two streams are interleaved into a single autoregressive sequence, enabling cross-stream interaction while preserving their distinct representations.
        
    \item We develop a source-aware neural audio codec with source-specific semantic distillation, producing semantically distinct speech and environment token streams from mixed audio. These streams provide explicit generation targets for the LLM and can be extracted from in-the-wild recordings without isolated sources.

    \item The framework allows the entire model to be trained directly on in-the-wild recordings. The resulting model outperforms baselines trained on the same simulated data in intelligibility, further improves scene relevance and audio quality through real-world adaptation, and preserves strong clean-speech generation.
    
    \item We demonstrate the effectiveness of distinct representations, interleaved generation, and real-world adaptation through ablations, and analyze whether the environmental sound remains stable while speech is active and how the two streams interact.
\end{itemize}
\section{Related Work}

\subsection{Speech-Scene Generation}
Speech-scene generation has been approached under two settings that differ in how the target scene is specified. One line of work takes the target scene from a reference recording~\cite{tan,ideatts,umbratts,daientts}, while the other, specifies it in natural language.
In the latter case, VoiceLDM~\cite{voiceldm} extends a text-to-audio latent diffusion model to follow both a scene description and speech content, with dual classifier-free guidance over the two conditions. VoiceDiT~\cite{voicedit} and ImmersiveTTS~\cite{immersivetts} instead obtain text-aligned speech features from a Glow-TTS-based~\cite{glowtts} speech module and generate the scene with a diffusion Transformer~\cite{dit}. VoiceDiT concatenates these features with the noisy latent, whereas ImmersiveTTS fuses them with the scene description through joint attention. ControlAudio~\cite{controlaudio}, developed for controllable text-to-audio generation, unifies text, timing, and phoneme information in a structured prompt. In all of these techniques, the acoustic target is a single latent representation of the mixed scene, whereas our framework generates speech and environmental audio as distinct token streams within a shared autoregressive context.

These techniques also differ in whether and how real-world recordings are used for training. VoiceDiT is pretrained on simulated scenes and then fine-tuned the pretrained model on AudioSet-Speech [11] with its speech module frozen. ImmersiveTTS is trained only on simulated scenes, since its representation alignment uses features of the clean speech before mixing. ControlAudio combines real-world recordings that have human-annotated timestamps with simulated scenes. In our framework, the codec extracts both generation targets directly from the mixed audio, so the entire model can be trained on in-the-wild recordings given only their transcripts and scene descriptions.

\subsection{Source-Aware Neural Audio Codecs}
Neural audio codecs quantize waveforms into sequences of discrete tokens, typically with residual vector quantization~(RVQ)~\cite{soundstream,encodec,dac} or a single quantizer in low-frame-rate designs~\cite{wavtokenizer}. Recent codecs distill semantic information from pretrained self-supervised models into the tokens to make them better suited for language modeling~\cite{speechtokenizer,moshi,xcodec,omnicodec}. 
Some codecs further represent the sources of a mixture separately, in the latent space~\cite{sanac,omran}, in distinct codebooks~\cite{sdcodec}, or in orthogonal subspaces~\cite{decodec}. 
However, these codecs are mainly designed for coding, separation, or enhancement of existing audio. Our codec shares the idea of source-specific quantization but is designed to provide generation targets for a language model. Source-specific semantic distillation makes each stream capture the semantics of its own source, so that the two streams extracted from a mixture serve as distinct, temporally aligned targets for joint scene generation.

\subsection{Language Modeling for Audio Generation}

Text-guided audio generation, including the speech-scene generation techniques above, has largely relied on diffusion or flow matching over continuous latents~\cite{audioldm,makeanaudio,tangoflux}. An alternative paradigm formulates generation as language modeling over discrete codec tokens, enabling autoregressive Transformers~\cite{transformer} to generate speech~\cite{valle}, music~\cite{musicgen}, and general audio~\cite{audiolm,audiogen,uniaudio}. Since a codec represents each frame with multiple tokens, some models delay or flatten the tokens along the time axis~\cite{musicgen}, while others predict them along the depth axis with a depth Transformer conditioned on a temporal backbone~\cite{rqtransformer,uniaudio,moshi,qwen3tts}. We follow the latter design with a pretrained LLM as the temporal backbone. To generate the speech and environment streams together, they are interleaved into a single autoregressive sequence, and a separate depth Transformer for each stream predicts the multiple tokens of each frame.

\begin{figure*}[t]
\centering
\includegraphics[width=\textwidth]{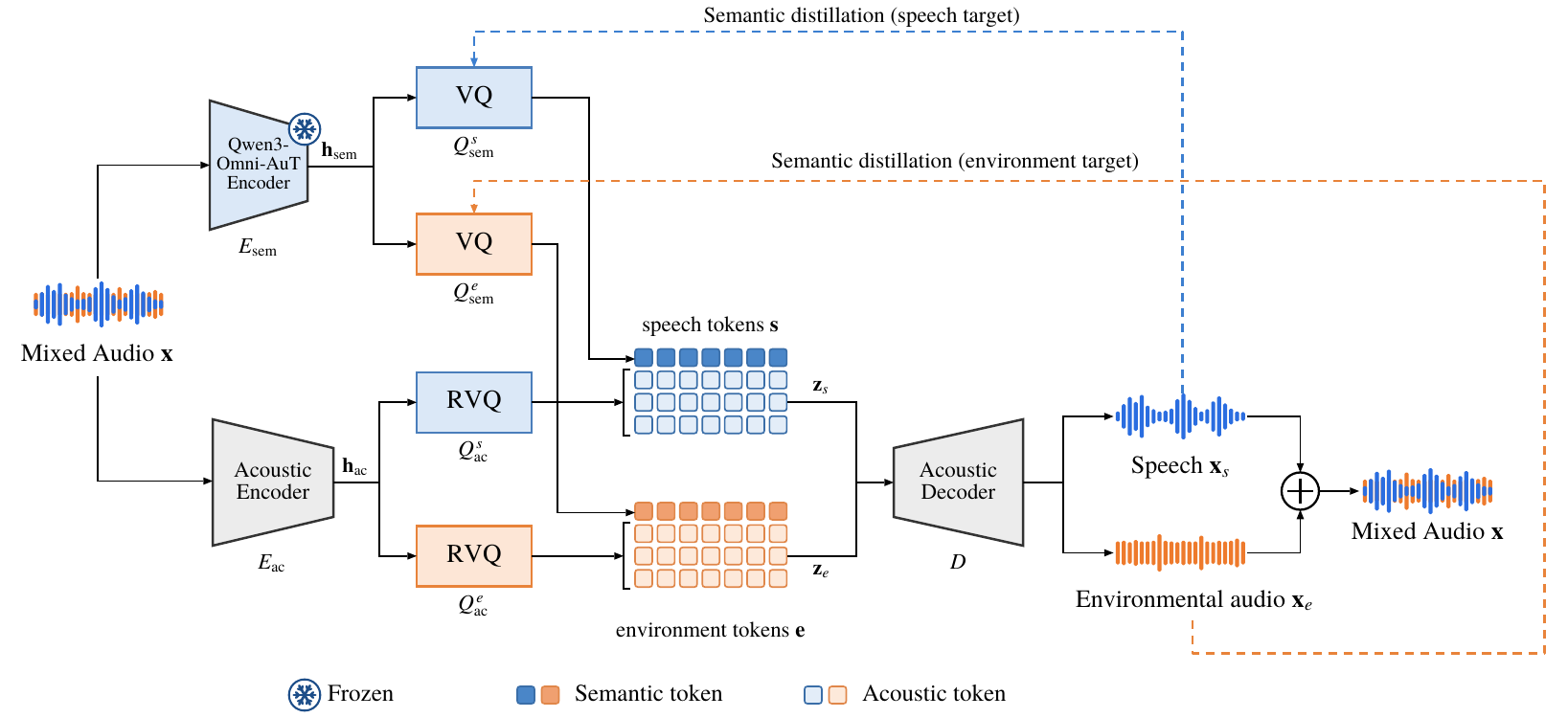}
\caption{Overview of the proposed source-aware neural audio codec. The semantic and acoustic representations extracted from the mixed waveform are shared by the two streams and quantized by source-specific VQ and RVQ modules into speech and environment tokens. Source-specific semantic distillation guides each VQ module with its own isolated source as the target, and the shared decoder reconstructs the speech and the environmental audio separately.}
\label{fig:codec}
\end{figure*}

\section{Proposed Method}
\subsection{Source-Aware Neural Audio Codec}
We construct a source-aware neural audio codec that converts a mixed speech scene into two temporally aligned discrete token streams representing speech and environmental audio, respectively. The codec builds on a backbone with separate semantic and acoustic branches and introduces source-specific quantization and supervision on top of it. It is trained on simulated scenes $\mathbf{x}=\mathbf{x}_s+\mathbf{x}_e$ mixed from separately recorded speech $\mathbf{x}_s$ and environmental audio $\mathbf{x}_e$, as illustrated in Fig.~\ref{fig:codec}.

\subsubsection{Codec Backbone}
We adopt OmniCodec~\cite{omnicodec} as the backbone, which consists of a semantic branch and an acoustic branch. The semantic branch extracts $\mathbf{h}_{\mathrm{sem}}=E_{\mathrm{sem}}(\mathbf{x})$ with the frozen audio encoder of Qwen3-Omni~\cite{qwen3omni} and quantizes it with a vector quantization (VQ) module. The acoustic branch extracts $\mathbf{h}_{\mathrm{ac}}=E_{\mathrm{ac}}(\mathbf{x})$ with a convolutional encoder, subtracts the quantized semantic representation from it, and quantizes the remainder with a residual vector quantization~(RVQ) module.  Each frame is represented by one semantic token and multiple acoustic tokens. The two quantized representations are summed into a latent $\mathbf{z}$ and decoded into a waveform $\hat{\mathbf{x}}=D(\mathbf{z})$. 
The backbone is trained with four groups of objectives. A distillation loss $\mathcal{L}_{\mathrm{dist}}$ aligns the quantized semantic representation with the features of a frozen self-supervised encoder $E_{\mathrm{ssl}}$, so that the semantic token captures the semantics of the input. A semantic reconstruction loss $\mathcal{L}_{\mathrm{sem}}$ recovers $\mathbf{h}_{\mathrm{sem}}$ from the quantized semantic representation, which prevents information loss in the VQ module. A reconstruction loss $\mathcal{L}_{\mathrm{rec}}$ applies a multi-scale mel-spectrogram loss and adversarial losses to the decoded waveform. Commitment and self-guidance losses $\mathcal{L}_{\mathrm{reg}}$ regularize the VQ and RVQ modules.

\subsubsection{Source-Specific Quantization}
Both semantic and acoustic branches encode the mixed waveform $\mathbf{x}$, so $\mathbf{h}_{\mathrm{sem}}$ and $\mathbf{h}_{\mathrm{ac}}$ are shared by the speech and environment streams. 
To obtain a separate token stream for each source, we use two semantic VQ modules, $Q^{s}_{\mathrm{sem}}$ for speech and $Q^{e}_{\mathrm{sem}}$ for environmental audio,
\begin{equation}
(\mathbf{q}^{s}_{\mathrm{sem}},\mathbf{s}^{(0)})=Q^{s}_{\mathrm{sem}}(\mathbf{h}_{\mathrm{sem}}),\quad
(\mathbf{q}^{e}_{\mathrm{sem}},\mathbf{e}^{(0)})=Q^{e}_{\mathrm{sem}}(\mathbf{h}_{\mathrm{sem}}),
\end{equation}
where $\mathbf{q}^{s}_{\mathrm{sem}}$ and $\mathbf{q}^{e}_{\mathrm{sem}}$ are the quantized semantic representations and $\mathbf{s}^{(0)}$ and $\mathbf{e}^{(0)}$ are the semantic token sequences. 
Likewise, we use two RVQ modules with $M$ layers, $Q^{s}_{\mathrm{ac}}$ for speech and $Q^{e}_{\mathrm{ac}}$ for environmental audio, each of which produces $M$ acoustic tokens per frame from $\mathbf{h}_{\mathrm{ac}}$. The tokens of the two streams at frame $t$ are
\begin{equation}
\mathbf{s}_t=(s^{(0)}_t,s^{(1)}_t,\ldots,s^{(M)}_t),\quad
\mathbf{e}_t=(e^{(0)}_t,e^{(1)}_t,\ldots,e^{(M)}_t),
\end{equation}
where the first element is the semantic token and the rest are the acoustic tokens. The two streams use disjoint codebooks and keep explicit source identities. 

\subsubsection{Source-Aware Training}
The codec is trained on simulated scenes, where the encoders take the mixture $\mathbf{x}$ and the isolated sources $\mathbf{x}_s$ and $\mathbf{x}_e$ serve only as targets. We keep the objectives of the backbone and apply them in a source-aware manner. The distillation loss is computed for each VQ module with the features of its own isolated source as the target,
\begin{equation}
\mathcal{L}_{\mathrm{dist}}=\ell_{\mathrm{dist}}\big(\mathbf{q}^{s}_{\mathrm{sem}},E_{\mathrm{ssl}}(\mathbf{x}_s)\big)+\ell_{\mathrm{dist}}\big(\mathbf{q}^{e}_{\mathrm{sem}},E_{\mathrm{ssl}}(\mathbf{x}_e)\big),
\end{equation}
where $\ell_{\mathrm{dist}}(\cdot,\cdot)$ is the distillation loss of the backbone. Although both VQ modules take the same $\mathbf{h}_{\mathrm{sem}}$ as input, the different distillation targets lead each to capture the semantics of its own source. The semantic reconstruction loss $\mathcal{L}_{\mathrm{sem}}$ likewise takes $E_{\mathrm{sem}}(\mathbf{x}_s)$ and $E_{\mathrm{sem}}(\mathbf{x}_e)$ as its targets.
For reconstruction, the quantized representations of each stream are summed into latents $\mathbf{z}_s$ and $\mathbf{z}_e$. Each latent is decoded by the shared decoder and supervised by its isolated source, and the mixture is reconstructed from the sum of the two latents,
\begin{equation}
\begin{aligned}
\mathcal{L}^{\mathrm{src}}_{\mathrm{rec}}&=\mathcal{L}_{\mathrm{rec}}\big(D(\mathbf{z}_s),\mathbf{x}_s\big)+\mathcal{L}_{\mathrm{rec}}\big(D(\mathbf{z}_e),\mathbf{x}_e\big),\\
\mathcal{L}^{\mathrm{mix}}_{\mathrm{rec}}&=\mathcal{L}_{\mathrm{rec}}\big(D(\mathbf{z}_s+\mathbf{z}_e),\mathbf{x}\big).
\end{aligned}
\end{equation}
The mixture-level term keeps the two streams complementary. The regularization losses $\mathcal{L}_{\mathrm{reg}}$ are applied to all VQ and RVQ modules of both streams. The overall objective is
\begin{equation}
\mathcal{L}_{\mathrm{codec}}=\mathcal{L}_{\mathrm{dist}}+\mathcal{L}_{\mathrm{sem}}+\mathcal{L}^{\mathrm{src}}_{\mathrm{rec}}+\mathcal{L}^{\mathrm{mix}}_{\mathrm{rec}}+\mathcal{L}_{\mathrm{reg}},
\end{equation}
with all loss weights following the backbone.

\subsubsection{Tokenization and Decoding}
Once trained, the codec produces both token streams directly from any mixed waveform,
\begin{equation}
(\mathbf{s},\mathbf{e})=\mathrm{CodecEnc}(\mathbf{x}),
\end{equation}
where $\mathbf{s}=(\mathbf{s}_1,\ldots,\mathbf{s}_T)$ and $\mathbf{e}=(\mathbf{e}_1,\ldots,\mathbf{e}_T)$ are the speech and environment token streams of a scene with $T$ frames. For waveform synthesis, the two token streams are mapped back to $\mathbf{z}_s$ and $\mathbf{z}_e$, each latent is decoded by the shared decoder, and the two waveforms are summed,
\begin{equation}
\hat{\mathbf{x}}=D(\mathbf{z}_s)+D(\mathbf{z}_e).
\end{equation}
Each stream can also be decoded on its own as $D(\mathbf{z}_s)$ or $D(\mathbf{z}_e)$.

\begin{figure*}[t]
\centering
\includegraphics[width=\textwidth]{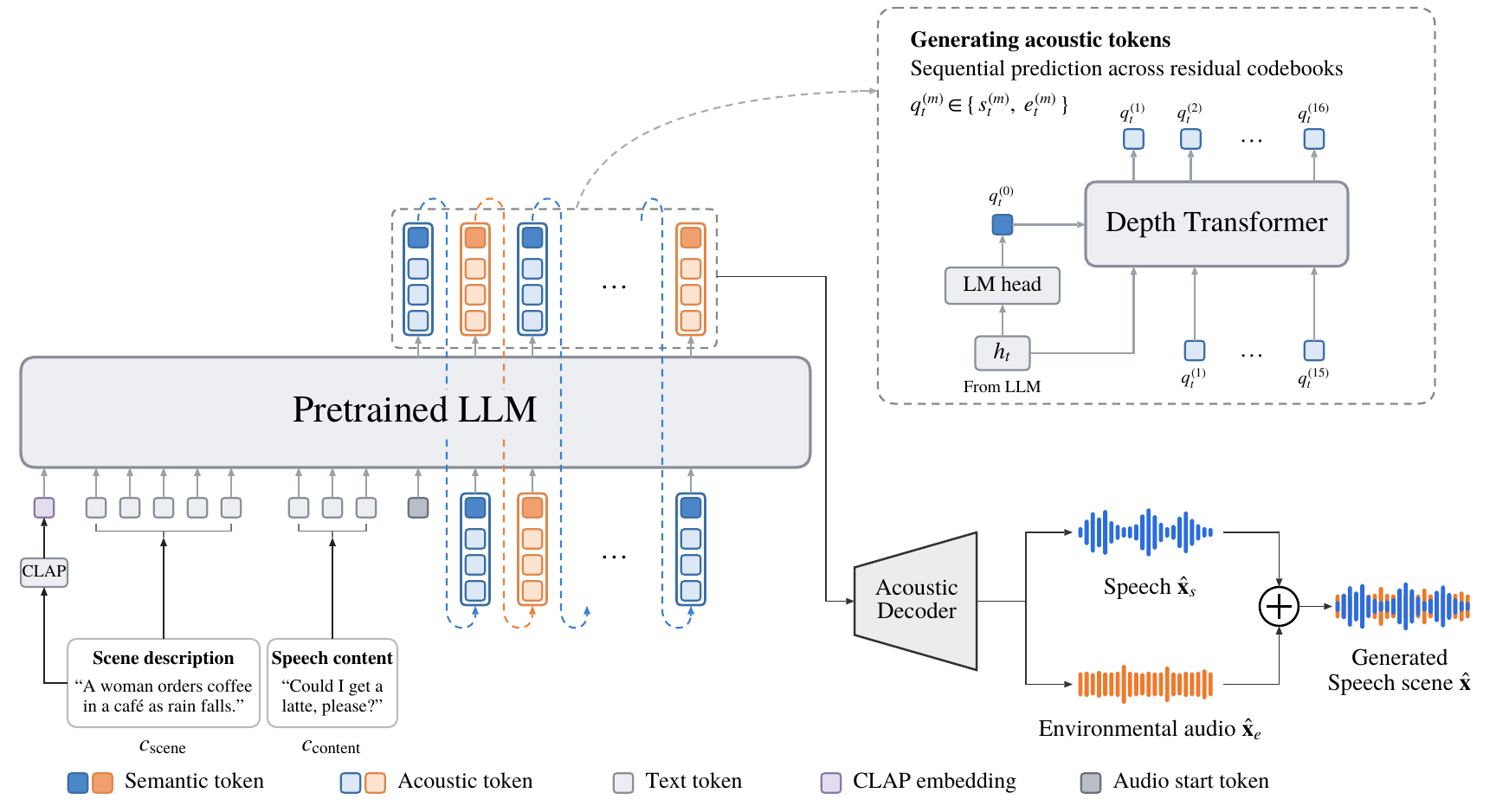}
\caption{Overview of the proposed LLM-based speech-scene generation framework. Given a scene description and target speech content, the pretrained LLM autoregressively generates interleaved speech and environment semantic tokens. For each generated semantic token, the corresponding acoustic tokens are predicted codebook-wise by a depth Transformer. The generated speech and environment token streams are then decoded into waveforms and combined to form the final speech scene.}
\label{fig:main_model}
\end{figure*}



\subsection{LLM-Based Speech-Scene Generation}
Given the speech content \(c_{\mathrm{content}}\) and the scene description \(c_{\mathrm{scene}}\), we formulate speech-scene generation as modeling the conditional distribution of the two token streams produced by the source-aware codec,
\begin{equation}
p(\mathbf{s}, \mathbf{e} \mid c_{\mathrm{content}}, c_{\mathrm{scene}}).
\end{equation}
We parameterize this distribution with a pretrained LLM extended to generate audio tokens, as illustrated in Fig.~\ref{fig:main_model}.
The scene description and the speech content are provided as text tokens, preceded by a CLAP embedding~\cite{clap} of the scene description as a supplementary prefix. Following a special audio start token that marks the beginning of the audio sequence, the LLM generates the speech and environment tokens autoregressively in an interleaved order. At each position, the LM head predicts the semantic token and a depth Transformer generates the acoustic tokens.

\subsubsection{Interleaved Speech--Environment Generation}
We model the speech and environment tokens as a single interleaved autoregressive sequence,
\begin{equation}
[\mathbf{s}_1,\mathbf{e}_1,\mathbf{s}_2,\mathbf{e}_2,\ldots,\mathbf{s}_T,\mathbf{e}_T].
\end{equation}
In this sequence, the tokens of the two streams for the same frame, $\mathbf{s}_t$ and $\mathbf{e}_t$, are placed next to each other, with $\mathbf{s}_t$ followed by $\mathbf{e}_t$. The interleaved ordering allows the speech and environment tokens to remain explicitly distinct while being modeled within the same autoregressive context. In particular, the environment tokens at frame $t$ are modeled with access to the corresponding speech tokens, while subsequent speech tokens can condition on the preceding environment tokens. The two streams continuously influence each other throughout the autoregressive sequence.

\subsubsection{Depth Transformer}
For each speech and environment semantic token generated by the LLM, a separate depth Transformer for each stream predicts the corresponding acoustic tokens. Specifically, after predicting $s^{(0)}_t$, the LLM hidden state $\mathbf{h}_{s,t}$ and the predicted semantic token are passed to the depth Transformer, which autoregressively generates the acoustic tokens along the codec depth dimension. This process is expressed as
\begin{equation}
p\big(s^{(1:M)}_t\mid s^{(0)}_t,\mathbf{h}_{s,t}\big)=\prod_{m=1}^{M}p\big(s^{(m)}_t\mid s^{(0:m-1)}_t,\mathbf{h}_{s,t}\big).
\end{equation}
Likewise, the environment acoustic tokens are generated as
\begin{equation}
p\big(e^{(1:M)}_t\mid e^{(0)}_t,\mathbf{h}_{e,t}\big)=\prod_{m=1}^{M}p\big(e^{(m)}_t\mid e^{(0:m-1)}_t,\mathbf{h}_{e,t}\big).
\end{equation}
After the acoustic tokens are generated, their embeddings are combined with that of the semantic token to form the input for the subsequent LLM step. The LLM context contains both the semantic and acoustic tokens of all previous frames, so that each prediction is conditioned on the full history of both streams.

The LLM and depth Transformer are optimized end-to-end using source-specific semantic and acoustic token prediction losses. Following the interleaved ordering defined above, the semantic token losses are
\begin{equation}
\begin{aligned}
\mathcal{L}_{s,\mathrm{sem}}&=-\sum_{t=1}^{T}\log p\big(s^{(0)}_t\mid\mathbf{s}_{<t},\mathbf{e}_{<t},c_{\mathrm{content}},c_{\mathrm{scene}}\big),\\
\mathcal{L}_{e,\mathrm{sem}}&=-\sum_{t=1}^{T}\log p\big(e^{(0)}_t\mid\mathbf{s}_{\le t},\mathbf{e}_{<t},c_{\mathrm{content}},c_{\mathrm{scene}}\big).
\end{aligned}
\end{equation}
The acoustic token losses are
\begin{equation}
\begin{aligned}
\mathcal{L}_{s,\mathrm{ac}}&=-\sum_{t=1}^{T}\sum_{m=1}^{M}\log p\big(s^{(m)}_t\mid s^{(0:m-1)}_t,\mathbf{h}_{s,t}\big),\\
\mathcal{L}_{e,\mathrm{ac}}&=-\sum_{t=1}^{T}\sum_{m=1}^{M}\log p\big(e^{(m)}_t\mid e^{(0:m-1)}_t,\mathbf{h}_{e,t}\big).
\end{aligned}
\end{equation}
The overall training objective is
\begin{equation}
\begin{aligned}
\mathcal{L}_{\mathrm{gen}}={}&\lambda_{s,\mathrm{sem}}\mathcal{L}_{s,\mathrm{sem}}+\lambda_{e,\mathrm{sem}}\mathcal{L}_{e,\mathrm{sem}}\\
&+\lambda_{s,\mathrm{ac}}\mathcal{L}_{s,\mathrm{ac}}+\lambda_{e,\mathrm{ac}}\mathcal{L}_{e,\mathrm{ac}}.
\end{aligned}
\end{equation}

\subsection{Training Strategy}
\label{train_stg}
The model is trained in two stages, pretraining on simulated scenes followed by fine-tuning on real-world recordings. The composition and spoken content of simulated scenes are known by construction, so pretraining establishes the role of each token stream and how the speech content and the scene description correspond to the two streams. Fine-tuning then exposes the model to the dependencies between speech and environment in real-world recordings, which mixing independent sources cannot reproduce. In both stages, the frozen codec is applied directly to the mixed waveform to obtain the two token streams, so the entire model is trained in the same way without isolated sources.

\subsubsection{Simulated Pretraining}
Existing pipelines~\cite{voicedit,immersivetts,controlaudio} for constructing simulated scenes draw speech from clean read-speech corpora, whose speaking style differs markedly from that of real-world recordings. We additionally use speech segmented from in-the-wild recordings after removing its background sounds by speech enhancement. These utterances retain the speaking style of real-world scenes, while the environment can still be chosen freely. Following~\cite{immersivetts}, the environmental audio is extended beyond the utterance, and the utterance is placed at a random position within it. Since the caption of the environmental recording describes only the environmental audio and carries no information about the speech, we collect speaker attributes such as gender for each utterance and compose the scene description from the environment caption and a description of the speaker, e.g., \textit{“A man speaks as rain falls on a hard surface."} or \textit{“Birds chirp in a park, and a woman talks."} The phrasing and the order of the two parts are randomly selected from a set of 140 templates so that the model does not rely on a fixed pattern and learns to ground each part in the corresponding stream.


\subsubsection{Real-World Adaptation}
Following simulated pretraining, we fine-tune the model on real-world recordings in which speech and environmental audio are captured together.
Real-world recordings, however, provide less reliable supervision than the simulated scenes. Their transcripts and scene descriptions are obtained automatically and may contain labeling errors, and we observe that prolonged adaptation degrades speech intelligibility. We keep the adaptation conservative by using an exponential moving average (EMA) of the model parameters. This allows the model to adapt to real-world speech scenes while limiting the loss of the linguistic capability acquired in pretraining.

\subsection{Classifier-Free Guidance}
During training, $c_{\mathrm{content}}$ and $c_{\mathrm{scene}}$ are each dropped independently with a small probability, which allows classifier-free guidance (CFG) to be applied to the two conditions separately at inference. At each generation step, the LLM runs three times, with both conditions, without $c_{\mathrm{scene}}$, and without $c_{\mathrm{content}}$. The guided hidden state is
\begin{multline}
\mathbf{h}_t^{\mathrm{cfg}}
= \mathbf{h}_t(c_{\mathrm{content}}, c_{\mathrm{scene}}) \\
+ \gamma_{\mathrm{content}}
  \big[\mathbf{h}_t(c_{\mathrm{content}}, c_{\mathrm{scene}})
     - \mathbf{h}_t(\varnothing, c_{\mathrm{scene}})\big] \\
+ \gamma_{\mathrm{scene}}
  \big[\mathbf{h}_t(c_{\mathrm{content}}, c_{\mathrm{scene}})
     - \mathbf{h}_t(c_{\mathrm{content}}, \varnothing)\big],
\end{multline}
where $\gamma_{\mathrm{content}}$ and $\gamma_{\mathrm{scene}}$ control the guidance strength toward the speech content and the scene, respectively. $\mathbf{h}_t^{\mathrm{cfg}}$ replaces $\mathbf{h}_{s,t}$ or $\mathbf{h}_{e,t}$ as the input to the LM head and the depth Transformer, so the semantic token and acoustic tokens at each position are generated from the same guided state. The two scales provide a controllable trade-off between speech intelligibility and scene relevance.

\section{Experimental Setup}

\subsection{Datasets}
\label{data}
\subsubsection{Simulated Speech Scenes}
The speech pool consists of 338K LibriTTS utterances and 815K utterances from an internal dataset derived from YODAS~\cite{yodas}. The YODAS recordings are processed with speech enhancement, speaker diarization, segmentation, transcription, and quality filtering, and speaker gender is annotated for every utterance in both pools. Environmental audio is drawn from 306K WavCaps~\cite{wavcaps} recordings, excluding those whose captions indicate foreground speech. Speech and environment are mixed at an SNR uniformly sampled between 2 and 10\,dB, with the utterance placed at a random position within the environment, and 15\% of the training samples contain speech only. Scene descriptions are composed from the WavCaps caption and the speaker gender as described in Sec.~\ref{train_stg}. The codec is trained on scenes mixed from the same speech and environmental audio but with the SNR sampled uniformly between 0 and 30~dB, covering a wider range of speech-to-environment levels.

\subsubsection{Real-World Speech Scenes}
We use English speech-containing recordings from AudioSet-Speech~\cite{voiceldm}. Following~\cite{voicedit}, we transcribe each recording with Whisper-large-v3~\cite{whisper} and crop it to its speech span. Scene descriptions are taken from the AudioSetCaps~\cite{audiosetcaps} captions, and recordings without a caption are excluded. We remove recordings containing singing or prominent music, and keep only those whose Whisper transcription achieves a WER of at most 15\% against the original transcript. The resulting training pool contains 269K recordings.

\subsubsection{Evaluation Set}
For speech-scene generation evaluation, we use the AC-filtered test set introduced by VoiceLDM~\cite{voiceldm}, which consists of speech-containing examples from the AudioCaps~\cite{audiocaps} test set with corresponding speech transcriptions. This benchmark has also been adopted by subsequent environment-aware speech generation systems~\cite{voicedit, immersivetts, controlaudio}. 

\subsection{Baselines and Compared Methods}
We compare the proposed model with publicly released or reproduced speech-scene generation systems using common captions, target transcripts, and evaluation implementations.

\textbf{VoiceLDM~\cite{voiceldm}.}
We use the official pretrained VoiceLDM-S and VoiceLDM-M checkpoints with the default classifier-free guidance settings. The two models share the same architecture and differ in the size of the U-Net backbone, following AudioLDM-S and AudioLDM-M, respectively. VoiceLDM is trained on real-world recordings from AudioSet~\cite{audioset}, CommonVoice~\cite{commonvoice}, and VoxCeleb1~\cite{voxceleb} together with non-speech audio from AudioSet and DEMAND~\cite{demand}, amounting to 2.43M speech and 824K non-speech segments, which is more than twice the amount of data used to train our model.

\textbf{ImmersiveTTS~\cite{immersivetts}.}
We train ImmersiveTTS using the official implementation on the same simulated speech scenes used to train our model. The model is trained for 400k steps, the same number of steps as in the original paper. Since its training requires separately available speech and environment components, ImmersiveTTS is trained only on simulated scenes. We search the content and environment guidance scales and obtain the best performance with scales of 3 and 3, which coincide with the default setting in [13].

\textbf{VoiceDiT~\cite{voicedit}.}
The training code and checkpoint of VoiceDiT are not publicly available, so it is not included in the objective comparison. It is evaluated only in the environment stability analysis in Sec.~\ref{env_stab} using samples collected from its official demo page.

\textbf{Proposed and Proposed-FT.}
Proposed denotes our model pretrained only on the simulated speech scenes, and is therefore directly comparable to ImmersiveTTS in training data. Proposed-FT denotes the same model after real-world adaptation on AudioSet-Speech. Guidance scales are denoted as $\gamma_{\mathrm{content}}/\gamma_{\mathrm{scene}}$. Proposed is evaluated with 1/3, and Proposed-FT with 1/3, 1/4, and 1/5 to examine the trade-off between speech intelligibility and scene relevance.

\subsection{Evaluation Metrics}

\textbf{Objective evaluation.}
We evaluate the generated speech scenes in terms of speech intelligibility, scene relevance, and audio quality. Speech intelligibility is measured by word error rate (WER) using English transcriptions from Whisper-large-v3. 
Scene relevance is measured by the cosine similarity between CLAP embeddings of the generated mixed waveform and its target scene caption, averaged over evaluation cases. The captions describe the complete acoustic scene and may include speech as well as environmental events.
Audio quality is evaluated using Fr\'echet Audio Distance (FAD) with VGGish~\cite{vggish} embeddings, which measures distributional similarity between generated and reference audios. Lower FAD indicates closer distributional agreement. For clean-speech evaluation, we additionally report UTMOS~\cite{utmos} as an automatic estimate of perceptual speech quality.

\textbf{Subjective evaluation.}
We conduct a mean opinion score (MOS) test on samples generated from the same AC-filtered ~\cite{voiceldm} evaluation set. For each model, 20 randomly selected samples are evaluated by 14 listeners using a five-point scale. We evaluate speech naturalness (MOS-SN), scene relevance (MOS-SR), and overall quality (MOS-OQ), and report the mean score with a 95\% confidence interval for each criterion.

\subsection{Implementation Details}
The generative model is initialized from Qwen3-0.6B-Base~\cite{qwen3}. Each depth Transformer has 5 layers, a hidden size of 1024, and 8 attention heads. During training, we use the CLAP text embedding of the scene description \(c_{\mathrm{scene}}\) with probability 0.8 and the CLAP audio embedding of the target audio with probability 0.2. At inference, the text embedding is always used. For classifier-free guidance, the scene description, its CLAP prefix, and the speech content \(c_{\mathrm{content}}\) are each dropped independently with probability 0.1.
The source-aware codec is initialized from the pretrained OmniCodec checkpoint~\footnote{\url{https://github.com/ASLP-lab/OmniCodec}} and trained for 50k steps with AdamW~\cite{adamw}. It operates on 24-kHz audio at 12.5 frames per second, and the VQ modules and all \(M=16\) layers of the RVQ modules use codebooks of size 2048. WavLM~\cite{wavlm} is used as the self-supervised encoder \(E_{\mathrm{ssl}}\) for semantic distillation.
Pretraining runs for 150k steps with Adam~\cite{adam}, a peak learning rate of \(2\times10^{-4}\), 20k warmup steps, and dynamic batches of up to 2,800 frames per GPU. We set \(\lambda_{\mathrm{s,sem}}=\lambda_{\mathrm{s,ac}}=1\) and \(\lambda_{\mathrm{e,sem}}=\lambda_{\mathrm{e,ac}}=0.5\), apply label smoothing of 0.05 to \(\mathcal{L}_{\mathrm{s,sem}}\) and \(\mathcal{L}_{\mathrm{e,sem}}\), and clip gradients at 5. Fine-tuning runs for 10k steps from the pretrained checkpoint with the same learning rate and batch configuration, 2.5k warmup steps, and an EMA decay of 0.9999. Decoding uses top-\(k\) sampling with \(k=50\) and a temperature of 0.9 for both semantic and acoustic tokens. All experiments use four NVIDIA RTX 4090 GPUs.

\section{Results and Discussion}

\begin{table}[t]
\centering
\caption{Objective evaluation of speech-scene generation.}
\label{tab:obj-eval}
\small
\setlength{\tabcolsep}{4pt}
\resizebox{\columnwidth}{!}{%
\begin{tabular}{lcccc}
\toprule
\textbf{Method}
& \textbf{Mix WER}($\downarrow$)
& \textbf{Speech WER}($\downarrow$)
& \textbf{CLAP}($\uparrow$)
& \textbf{FAD}($\downarrow$) \\
\midrule
GT & 20.91 & -- & 0.356 & -- \\
\midrule
VoiceLDM-S   & 46.90 & -- & 0.235 & 4.14 \\
VoiceLDM-M   & 8.44  & -- & 0.210 & 5.27 \\
ImmersiveTTS & 7.91  & -- & 0.247 & 4.03 \\
\midrule
\multicolumn{5}{l}{\textit{Simulated pretraining}} \\
Proposed & 6.43 & 5.26 & 0.255 & 4.76 \\
\midrule
\multicolumn{5}{l}{\textit{Real-world adaptation}} \\
Proposed-FT (CFG 1/3) & \textbf{6.27} & \textbf{5.25} & 0.264 & 4.17 \\
Proposed-FT (CFG 1/4) & 7.08 & 5.63 & 0.286 & 3.91 \\
Proposed-FT (CFG 1/5) & 7.31 & 6.10 & \textbf{0.293} & \textbf{3.71} \\
\bottomrule
\end{tabular}%
}
\end{table}

\subsection{Main Results}
Table~\ref{tab:obj-eval} reports the main speech-scene generation results on the AC-filtered evaluation set. We compare the proposed model with the baselines in terms of speech intelligibility, scene relevance, and audio quality. 
For the proposed models, we additionally report \textit{Speech WER}, computed on the independently decoded speech stream, alongside \textit{Mix WER} measured on the final mixed waveform. 
The high WER of the ground truth reflects that the reference recordings are naturally noisy and their transcripts are obtained automatically, so WER here measures intelligibility relative to the target content rather than exact reproduction of the reference.

Even when trained on the same simulated scenes, the proposed model outperforms ImmersiveTTS in both intelligibility and scene relevance, reducing WER from 7.91\% to 6.43\% and raising CLAP similarity from 0.247 to 0.255. Real-world adaptation then brings a clear improvement. As $\gamma_{\mathrm{scene}}$ increases from 3 to 5, CLAP rises from 0.264 to 0.293 and FAD drops from 4.17 to 3.71, showing that the generated scenes become closer to real-world recordings both in content and in distribution. WER increases moderately from 6.27\% to 7.31\%, which reflects a controllable trade-off between content fidelity and scene emphasis. Nevertheless, every setting of Proposed-FT remains more intelligible than all baselines, and the 1/4 and 1/5 settings also achieve the lowest FAD. Notably, Speech WER remains consistently lower than Mix WER across all proposed settings, showing that the explicit speech stream preserves highly intelligible linguistic content even when generating a complete speech scene. 


Table~\ref{tab:sub-eval} reports the subjective evaluation on the same evaluation set. Proposed-FT uses the CFG setting of 1/4. Proposed is rated clearly above both baselines on all three criteria. Proposed-FT improves on all three further, particularly in MOS-SN and MOS-SR, which confirms the effect of real-world adaptation. Between the two baselines, MOS-SN is similar, whereas ImmersiveTTS is rated higher in MOS-SR and VoiceLDM-M in MOS-OQ. ImmersiveTTS, with its environment-specific conditioning and alignment, follows the scene description more closely, but VoiceLDM-M, trained on real-world recordings, is rated higher in overall quality. The proposed model is preferred on every criterion, as it learns each component well from simulated scenes and then adapts to real-world recordings.

\subsection{Clean Speech Generation}
Table~\ref{tab:clean_speech_generation} evaluates clean-speech generation on LibriTTS~\cite{libritts} and LibriSpeech~\cite{librispeech}. For VoiceLDM and ImmersiveTTS, we follow the clean-speech generation settings reported in their original papers, including the corresponding classifier-free guidance configurations and scene descriptions. For the proposed model, we use CFG scales of 1/1 and provide a simple gender-dependent scene description, \textit{“A man is talking."} or \textit{“A woman is talking."}
Although the proposed model is primarily trained for speech-scene generation, it preserves strong clean-speech synthesis capability across all evaluation subsets. The pretrained model already achieves competitive WER, while showing a particularly large improvement in UTMOS over VoiceLDM-M and ImmersiveTTS. Real-world adaptation further improves intelligibility, reducing WER from 4.24 to 3.84 on LibriTTS test-clean and from 4.40 to 4.14 on LibriSpeech test-clean. Across the four evaluation subsets, Proposed-FT achieves the lowest WER among the compared generative systems in all cases and consistently outperforms ImmersiveTTS in speech naturalness.
We attribute this behavior in part to the use of a distinct speech token stream. In the proposed formulation, clean-speech generation and speech-scene generation share the same speech representation, with the main difference being whether an environment stream is generated alongside it. Consequently, learning speech within a mixed acoustic scene remains closely aligned with learning speech itself, rather than requiring linguistic and environmental information to be modeled through a single mixed target representation. This reduced discrepancy between the two generation settings may contribute to the strong transfer to clean-speech synthesis, particularly in terms of perceptual naturalness.

\begin{table}[t]
\centering
\caption{Subjective evaluation of speech-scene generation.}
\label{tab:sub-eval}
\small
\setlength{\tabcolsep}{6pt}
\begin{tabular}{lccc}
\toprule
\textbf{Method}
& \textbf{MOS-SN($\uparrow$)}
& \textbf{MOS-SR($\uparrow$)}
& \textbf{MOS-OQ($\uparrow$)} \\
\midrule
GT
& $4.28 \pm 0.13$
& $4.37 \pm 0.11$
& $4.34 \pm 0.11$ \\
\midrule
VoiceLDM-M
& $3.21 \pm 0.14$
& $2.99 \pm 0.14$
& $3.25 \pm 0.13$ \\
ImmersiveTTS
& $3.24 \pm 0.13$
& $3.09 \pm 0.14$
& $3.07 \pm 0.12$ \\
\midrule
Proposed
& $3.44 \pm 0.11$
& $3.39 \pm 0.13$
& $3.43 \pm 0.12$ \\
Proposed-FT
& $\mathbf{3.57 \pm 0.11}$
& $\mathbf{3.53 \pm 0.13}$
& $\mathbf{3.48 \pm 0.12}$ \\
\bottomrule
\end{tabular}
\end{table}

\begin{table}[!h]
\caption{Reconstruction Quality on AudioSet-Speech.}
\label{tab:codec_recon}
\centering
\setlength{\tabcolsep}{4.5pt}
\resizebox{\columnwidth}{!}{%
\begin{tabular}{lcccc}
\toprule
\textbf{Codec} & \textbf{WER}($\downarrow$) & \textbf{CLAP}($\uparrow$) & \textbf{FAD}($\downarrow$) & \textbf{UTMOS}($\uparrow$) \\
\midrule
GT                  & 1.24 & 0.202 & --    & 2.23 \\
\midrule
OmniCodec           & \textbf{3.85} & 0.192 & \textbf{0.533} & 2.25 \\
Source-aware codec & 4.94 & \textbf{0.209} & 0.562 & \textbf{2.40} \\
\bottomrule
\end{tabular}%
}
\end{table}

\begin{table}[!h]
\caption{Effect of Distinct Source-aware Token Representation.}
\label{tab:representation_ablation}
\centering
\setlength{\tabcolsep}{4.5pt}
\resizebox{\columnwidth}{!}{%
\begin{tabular}{llccc}
\toprule
\textbf{Stage} & \textbf{Representation} & \textbf{WER}($\downarrow$) & \textbf{CLAP}($\uparrow$) & \textbf{FAD}($\downarrow$) \\
\midrule
\multirow{2}{*}{\textit{Simulated pretraining}}
& Mixed      & 11.33 & 0.235 & 6.66 \\
& Distinct   & \textbf{6.43} & \textbf{0.255} & \textbf{4.76} \\
\midrule
\multirow{2}{*}{\textit{Real-world adaptation}}
& Mixed      & 8.76 & 0.239 & 6.30 \\
& Distinct   & \textbf{6.27} & \textbf{0.264} & \textbf{4.17} \\
\bottomrule
\end{tabular}%
}
\end{table}

\begin{table*}[t]
\centering
\caption{Comparison of clean speech generation performance on LibriTTS and LibriSpeech.}
\label{tab:clean_speech_generation}
\small
\setlength{\tabcolsep}{4pt}
\begin{tabular}{lcccccccc}
\toprule
\multirow{3}{*}{\textbf{Method}}
& \multicolumn{4}{c}{\textbf{LibriTTS}}
& \multicolumn{4}{c}{\textbf{LibriSpeech}} \\
\cmidrule(lr){2-5}
\cmidrule(lr){6-9}
& \multicolumn{2}{c}{\textbf{test-clean}}
& \multicolumn{2}{c}{\textbf{test-other}}
& \multicolumn{2}{c}{\textbf{test-clean}}
& \multicolumn{2}{c}{\textbf{test-other}} \\
\cmidrule(lr){2-3}
\cmidrule(lr){4-5}
\cmidrule(lr){6-7}
\cmidrule(lr){8-9}
& \textbf{WER}($\downarrow$) & \textbf{UTMOS}($\uparrow$)
& \textbf{WER}($\downarrow$) & \textbf{UTMOS}($\uparrow$)
& \textbf{WER}($\downarrow$) & \textbf{UTMOS}($\uparrow$)
& \textbf{WER}($\downarrow$) & \textbf{UTMOS}($\uparrow$) \\
\midrule
GT
& 3.47 & 4.10
& 5.13 & 3.53
& 2.89 & 4.12
& 4.95 & 3.56 \\
\midrule
VoiceLDM-M
& 4.25 & 3.11
& 4.52 & 3.11
& 5.21 & 3.01
& 4.90 & 3.01 \\
ImmersiveTTS
& 5.03 & 3.06
& 4.59 & 3.05
& 6.07 & 3.04
& 6.50 & 3.00 \\
\midrule
Proposed
& 4.24 & \textbf{3.69}
& 4.65 & \textbf{3.65}
& 4.40 & \textbf{3.64}
& 4.98 & 3.61 \\
Proposed-FT
& \textbf{3.84} & \textbf{3.69}
& \textbf{4.21} & 3.64
& \textbf{4.14} & \textbf{3.64}
& \textbf{4.84} & \textbf{3.62} \\
\bottomrule
\end{tabular}
\end{table*}






\subsection{Ablation: Distinct Representation}
We examine whether distinct speech and environment token streams serve as a more suitable generation target for speech-scene generation than a single mixed token stream. As the mixed-token baseline, we use the pretrained OmniCodec backbone as is, which encodes the entire scene into a single token stream, and train the same generative model on it with identical data, optimization, and decoding settings. Since the baseline has a single token stream, it uses one depth Transformer, which predicts all 32 RVQ layers of the original OmniCodec, so that the number of tokens per frame is comparable to that of the two streams combined in the proposed model. Both models are evaluated with the CFG setting of 1/3 at both stages. Note that this baseline is not disadvantaged at the codec level. Table~\ref{tab:codec_recon} compares the reconstruction quality of the two codecs on 1,000 recordings randomly sampled from the AudioSet-Speech pool. On these recordings, the original OmniCodec outperforms our source-aware codec in both FAD and WER.

However, in generation, Table~\ref{tab:representation_ablation} shows that the distinct representation is better on every metric at both stages. After simulated pretraining, the distinct model achieves 6.43\% WER, 0.255 CLAP, and 4.76 FAD, compared with 11.33\%, 0.235, and 6.66 for the mixed baseline. After real-world adaptation, the mixed baseline improves to 8.76\% WER but remains well behind the distinct model, and the gaps in CLAP and FAD stay large. 
We attribute this to the structure of the target, in which each source has its own prediction space instead of sharing the same tokens. The gaps in WER and CLAP confirm that preserving the identities of speech and environment in the generation target is essential for learning both components well, even at the cost of lower reconstruction fidelity.

\subsection{Ablation: Interleaved Versus Parallel Generation}
We next examine how the two streams should be ordered in the autoregressive sequence. The proposed model interleaves them so that $\mathbf{s}_t$ and $\mathbf{e}_t$ occupy consecutive positions, as in~(9). The parallel variant instead predicts both streams at the same step from a single hidden state through two output heads, so that the two tokens of a frame share the preceding context but cannot condition on each other within the frame. All other components are identical.

Table~\ref{tab:interleave_ablation} shows a clear and consistent difference. The interleaved model yields a much lower WER at every stage and guidance setting, whereas the parallel model obtains slightly higher CLAP and lower FAD after adaptation. 
The two orderings favor different components. We conjecture that, in parallel prediction, a single hidden state must simultaneously support two heterogeneous targets with different prediction characteristics. This may introduce interference between the speech and environment streams, particularly for the more structured speech target. Interleaving instead assigns a separate autoregressive position and hidden state to each stream, while allowing the environment token at each frame to condition on the corresponding speech tokens. This substantially improves WER, while the parallel ordering is more favorable in terms of scene relevance and audio quality. Since intelligible speech is a primary requirement of speech-scene generation, we adopt the interleaved ordering.

\begin{table}[!h]
\caption{Comparison of Interleaved and Parallel Generation.}
\label{tab:interleave_ablation}
\centering
\setlength{\tabcolsep}{4.5pt}
\resizebox{\columnwidth}{!}{%
\begin{tabular}{llccccc}
\toprule
\multirow{2}{*}{\textbf{Stage}} &
\multirow{2}{*}{\textbf{Generation}} &
\multicolumn{2}{c}{\textbf{CFG}} &
\multirow{2}{*}{\textbf{WER}($\downarrow$)} &
\multirow{2}{*}{\textbf{CLAP}($\uparrow$)} &
\multirow{2}{*}{\textbf{FAD}($\downarrow$)} \\
\cmidrule(lr){3-4}
& & \textbf{Content} & \textbf{Scene} & & & \\
\midrule
\multirow{2}{*}{\textit{Simulated pretraining}}
& Parallel    & 1 & 3 & 10.41 & 0.248 & \textbf{4.11} \\
& Interleaved & 1 & 3 & \textbf{6.43} & \textbf{0.255} & 4.76 \\
\midrule
\multirow{6}{*}[-0.8ex]{\textit{Real-world adaptation}}
& Parallel    & 1 & 3 & 9.68  & \textbf{0.271} & \textbf{3.41} \\
& Interleaved & 1 & 3 & \textbf{6.27} & 0.264 & 4.17 \\
\cmidrule(lr){2-7}
& Parallel    & 1 & 4 & 9.39  & \textbf{0.292} & \textbf{3.23} \\
& Interleaved & 1 & 4 & \textbf{7.08} & 0.286 & 3.91 \\
\cmidrule(lr){2-7}
& Parallel    & 1 & 5 & 10.87 & \textbf{0.296} & \textbf{3.07} \\
& Interleaved & 1 & 5 & \textbf{7.31} & 0.293 & 3.71 \\
\bottomrule
\end{tabular}%
}
\end{table}

\subsection{Analysis: Environment Stability During Speech}
\label{env_stab}
We further analyze how the environment behaves while speech is active. In scenes with continuous environmental sounds such as rain or flowing water, the environment should remain stable regardless of speech activity. We observe, however, that some systems exhibit an \textit{unintentional ducking} effect, in which the environment is attenuated during speech and recovers after it ends.

To quantify this, we separate each generated scene into speech and environment components with the generalized BandSplit network~\cite{bandsplit} and detect speech-active regions on the separated speech with Silero VAD.~\footnote{https://huggingface.co/AXERA-TECH/SileroVAD} The same separator is applied to all systems, including ours, although our codec could decode the two streams separately. For each sample, the environment loudness is expressed relative to its average level in the non-speech region, which is set to 0 dB. The resulting curves are aligned at speech onset and at speech offset and averaged over samples, so that they show how the environmental loudness changes as speech begins and ends. For models with public checkpoints, we generate 3 samples per scene with different seeds and show the spread as a shaded range. VoiceDiT is evaluated on samples from its official demo page, and only around speech offset, since those samples place speech at the beginning of the scene. 

\begin{figure}[t]
\centering
\includegraphics[width=\columnwidth]{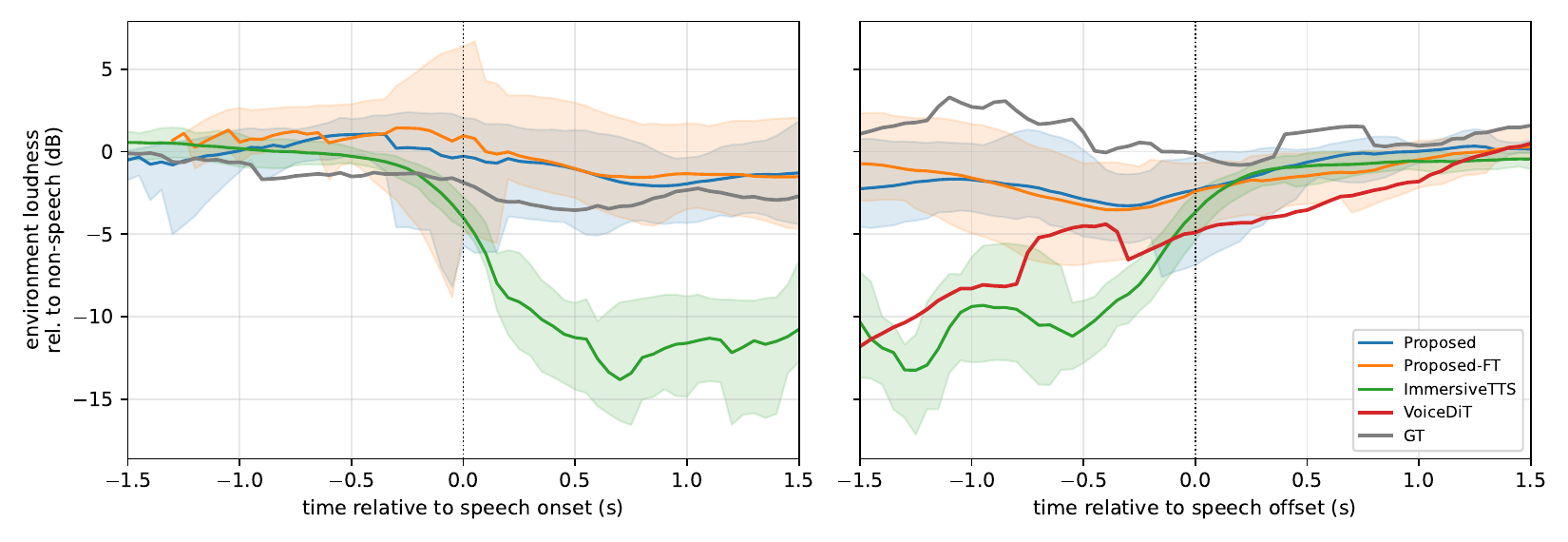}
\caption{Environmental loudness around speech activity for different speech-scene generation models. Loudness is measured from the separated environmental signal and normalized by the average level in non-speech regions.}
\label{fig:ducking}
\end{figure}

As shown in Fig.~\ref{fig:ducking}, ImmersiveTTS shows a pronounced drop in environmental loudness after speech onset and recovers toward the non-speech level around offset, and VoiceDiT shows a clear rise after speech ends, indicating suppression during speech. 
Both variants of the proposed model keep the environment close to the non-speech level throughout the speech region, closely following the ground truth. A plausible cause is that these systems inject frame-level speech conditioning through dedicated speech pathways into a single mixed target, so that speech activity modulates the environment as a side effect. Keeping the two streams distinct reduces this coupling and lets continuous environmental sounds remain stable during speech.

\subsection{Analysis: Interaction Between Speech and Environment Representations}

\begin{figure}[t]
\centering
\includegraphics[width=\columnwidth]{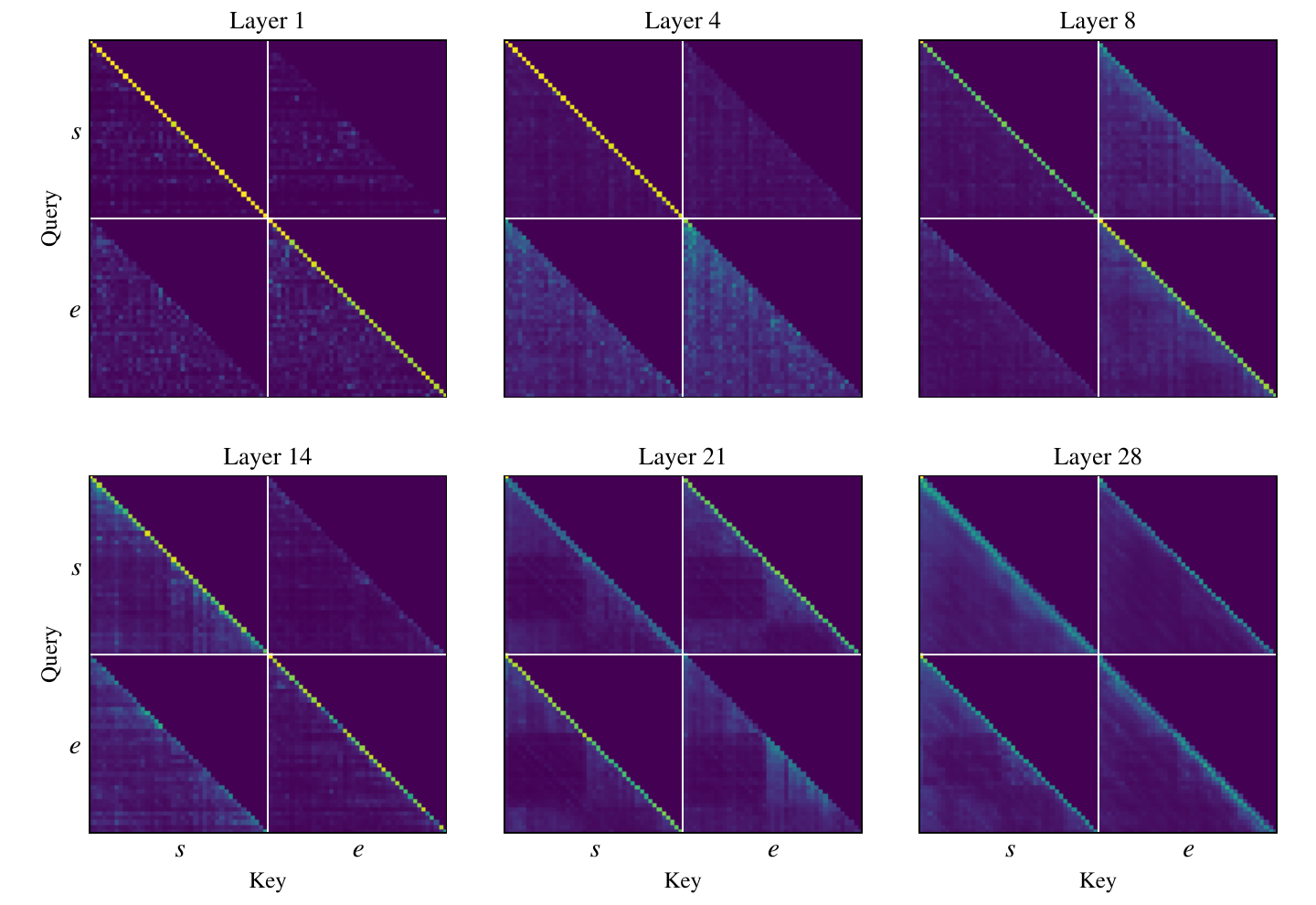}
\caption{Head-averaged self-attention of Proposed-FT at six layers for a generated scene. Positions are reordered so that speech tokens ($s$) precede environment tokens ($e$), yielding within-stream blocks on the diagonal and cross-stream blocks off the diagonal. Brighter colors indicate higher attention weight.}
\label{fig:attn}
\end{figure}

The interleaved formulation is intended to let the two streams condition on each other rather than to generate them independently. To see whether the model actually uses this, we inspect the self-attention of the LLM in Proposed-FT. For a generated scene, we average the attention weights over heads at each layer and group the positions by stream, which yields within-stream blocks ($s\!\to\!s$, $e\!\to\!e$) on the diagonal and cross-stream blocks ($s\!\to\!e$, $e\!\to\!s$) off the diagonal. Fig.~\ref{fig:attn} shows six of the 28 layers.

Three patterns stand out. First, the cross-stream blocks are far from empty: both $e\!\to\!s$ and $s\!\to\!e$ receive clear attention mass in most layers, so the streams are not treated as independent sequences. Second, the direction of interaction changes with depth. In the early layers, attention is dominated by the within-stream diagonal, with the environment attending to the speech stream more than the reverse, which is consistent with the ordering in which $\mathbf{e}_t$ is generated after $\mathbf{s}_t$. In the middle layers, the speech queries also attend to the environment tokens of nearby frames, so the conditioning becomes bidirectional. Third, in the deeper layers, all four blocks show a band along the diagonal, indicating that each stream attends mainly to the recent context of both streams, and the block-like structure within the bands suggests that attention is grouped over segments of similar acoustic content rather than individual frames.


\section{Limitations}

Real-world adaptation improves scene relevance and audio quality, but the AudioSet recordings used in this study are not specifically curated for speech-scene generation and rely on automatically obtained transcripts and scene descriptions.
The source-aware codec is also trained on a smaller and less diverse corpus than large general-purpose neural audio codecs. Scaling its training data, together with using better-curated real-world speech scenes, could further improve both the learned token representations and downstream generation quality.

\section{Conclusion}

We presented an LLM-based speech-scene generation framework that models speech and environmental audio as distinct but interacting token streams. A source-aware codec extracts temporally aligned speech and environment representations from mixed audio, while interleaved autoregressive generation allows the two streams to condition on each other within a shared context. The same representation also enables adaptation to in-the-wild recordings without requiring isolated sources.
We demonstrate that distinct representations improve both speech intelligibility and environmental modeling over a single mixed target, while interleaved generation further benefits speech intelligibility. Real-world adaptation improves scene relevance and audio quality, and additional analyses show stable environmental generation and clear cross-stream interaction. These results support explicitly preserving the different structures of speech and environment while jointly modeling their relationship for coherent speech-scene generation.

\bibliographystyle{IEEEtran}
\bibliography{ref}

\vfill

\end{document}